\documentclass[iop]{emulateapj-rtx4}

\shortauthors{Chauvin et al.}
\shorttitle{Hard X-ray polarimetry with \textit{INTEGRAL} SPI}

\begin{document}

\title{POLARIMETRY IN THE HARD X-RAY DOMAIN WITH \textit{INTEGRAL} SPI \footnote{
Based on observations with \textit{INTEGRAL}, an ESA project with instruments and science data center
funded by ESA member states (especially the PI countries: Denmark, France, Germany, Italy, 
Spain, and Switzerland), Czech Republic, and Poland with participation of Russia and USA.} }

\author{M. Chauvin\altaffilmark{1,2}, J. P. Roques\altaffilmark{1,2}, D. J. Clark\altaffilmark{3}, and E. Jourdain\altaffilmark{1,2}}
\altaffiltext{1}{Universit\'e de Toulouse, UPS-OMP, IRAP,  Toulouse, France;}
\altaffiltext{2}{CNRS, IRAP, 9 Av. Colonel Roche, BP 44346, F-31028 Toulouse Cedex 4, France} 
\altaffiltext{3}{Createc Ltd, Unit 8, Derwent Mill Commercial Park, Cockermouth, Cumbria CA13 0HT, UK}

\begin{abstract}
We present recent improvements in polarization analysis with the \textit{INTEGRAL} SPI data.
The SPI detector plane consists of 19 independent Ge crystals and can operate as a polarimeter. 
The anisotropy characteristics of Compton diffusions can provide information 
on the polarization parameters of the incident flux.
By including the physics of the polarized Compton process in the instrument simulation, 
we are able to determine the instrument response for a linearly polarized emission 
at any position angle. We compare the observed data with the 
simulation sets by a minimum $\chi^2$ technique to determine the polarization parameters 
of the source (angle and fraction).
We have tested our analysis procedure with Crab nebula observations and find a position 
angle similar to those previously reported in the literature, with a comfortable significance.
Since the instrument response depends on the incident angle, each exposure in the SPI data  
requires its own set of simulations, calculated for 18 polarization angles 
(from 0$^\circ$ to 170$^\circ$ in steps of 10$^\circ$) and unpolarized  emission. The 
analysis of a large amount of observations for a given source, required to obtain 
statistically significant results, represents a large amount of computing time, but it is 
the only way to access this complementary information in the hard X-ray regime. 
Indeed, major scientific advances are expected from such studies since the observational 
results will help to discriminate between the different models proposed for the high energy 
emission of compact objects like X-ray binaries and active galactic nuclei or gamma-ray bursts.
\end{abstract}      


\maketitle

\section{INTRODUCTION}
Throughout the universe we observe powerful engines which accelerate particles to immense 
energies. The precise details of how these engines function are still poorly understood, 
but polarization measurements of the high energy radiation are pivotal to gain a deeper 
insight into the physical environment in these systems.\\
Highly ordered geometries are generally required to produce 
a net polarization signal in the emission from a source, and strong magnetic fields are 
usually thought to play a leading role in the emission processes. The strength and direction of a 
polarization signal will lead back to an understanding of the physical mechanisms and their 
geometry at the emission site. There have been few attempts at measuring hard X-ray 
polarization since measurements are hampered by large backgrounds and systematic effects 
within the detectors. However, the advent of modern fast computing clusters has made large 
scale simulation of an instrument's response now possible. By combining instrument data and 
results from detailed Monte-Carlo Mass-Model simulations using GEANT4, it is possible to put 
constraints on the polarization characteristics of the hard X-ray flux emitted by a source. 
Using the Compton scattered events in the SPI instrument on \textit{INTEGRAL} (130-8000 keV), 
constraints have been already put on the percentage polarization in the prompt gamma-ray 
flux of gamma-ray bursts (GRBs; \citealt{2007A&A...466..895M}) and the Crab pulsar \citep{2008Sci...321.1183D}. 
The latter showing a remarkable alignment to the rotational axis of the spinning Neutron star.\\
However, improvements on the methods and techniques employed in the search for 
polarization could be achieved.
This paper discusses new developments that have been made in the mass-model and in the 
analysis procedure for the SPI instrument.

\section{DETECTION OF POLARIZED EMISSION}
\begin{figure}
\plotone{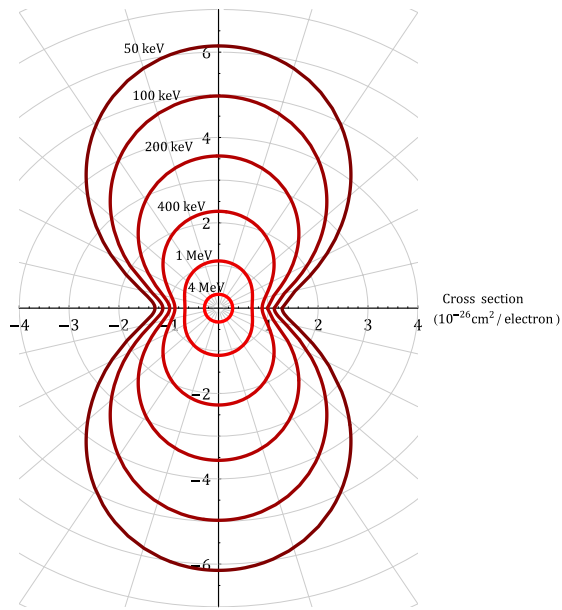}
\caption{Klein-Nishina differential cross-section for a photon polarized parallel to the $x$-axis, 
seen from the direction of incidence, showing the anisotropy in the azimuthal scatter direction. 
As the energy increases the photon is more likely to be forward scattered and the distribution 
becomes more isotropic \citep{2007ClarkThesis}.}
\label{KN_plot}
\end{figure}

The signature of polarization in the source signal can be found through the interaction 
process of photons within detectors.\\ 
In the hard X-ray regime, the dominant process is Compton scattering. In the case of a linearly 
polarized flux, the azimuthal angle $\phi$ of the scattered photon is no longer 
isotropic. The probability for the photon to be 
scattered at a polar angle $\theta$ and an azimuthal angle $\phi$ is given by the Klein-Nishina 
differential cross-section
\begin{equation}
\label{KN_equ}
\frac{d\sigma_\mathrm{KN,P}}{d\Omega}=\frac{1}{2}r_0^2\epsilon^2 \left[ \epsilon+\epsilon^{-1}-\sin^2\theta\cos^2\phi \right],
\end{equation}
where $r_0$ is the classical electron radius, $\epsilon$ is the ratio between the scattered 
and incident photon energies, and $\phi$ is the azimuthal scatter angle defined as the angle 
to the polarization unit vector. An azimuthal anisotropy is 
observed in the scattering direction, producing a greater number of scatters in a direction 
perpendicular to the emission polarization angle (P.A.; see Figure \ref{KN_plot}). This formula shows 
that as the energy increases the angular distribution related to the polarization becomes more 
isotropic.\\
Using a suitable arrangement of detector elements, this anisotropy can be used to determine 
the direction and degree of polarization of any observed emission.

\section{SPI AS A POLARIMETER}
\begin{figure}
\plotone{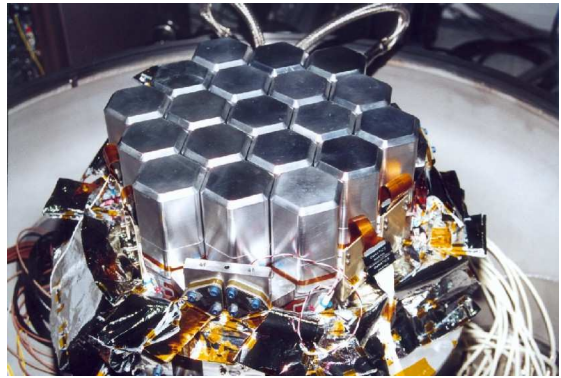}
\caption{The 19 germanium detectors before installation into the spacecraft (Credit: ESA).}
\label{Ge}
\end{figure}

SPI is a coded-aperture telescope on board the \textit{INTEGRAL} observatory operating between 
20 keV and 8 MeV. The description of the instrument and its performances can be found in 
\cite{2003A&A...411L..63V} and \cite{2003A&A...411L..91R}. SPI consists of 19 hexagonal Germanium 
detectors measuring 5.6 cm flat-to-flat and 7 cm deep, tessellated into a hexagonal shape 
(see Figure \ref{Ge}). The center-to-center distance between detectors is 6 cm. Originally designed 
for fine spectroscopy, SPI can be used as a Compton polarimeter. The measurement of polarization 
using Compton events relies on the positions of the initial and final photon interaction in the 
detectors. In SPI, there is no positional information available inside a detector so only the multiple 
events where the photon deposits its energy in more than one detector can be used. This kind of event 
is identified by energy deposits occurring within a coincidence window of 350 ns and not detected by 
the anti-coincidence system. The effective area of SPI for multiple events is reported in 
\cite{2003A&A...411L..71A} and range from 10 cm$^2$ to 50 cm$^2$ between 150 keV and 1 MeV. Since 
\textit{INTEGRAL}s launch, four detectors have failed: detector 2 (2003), detector 17 (2004), detector 5 (2009), 
and detector 1 (2010). These failures have resulted in a decrease of the effective area of the 
instrument down to $\sim52\%$ of the original area for multiple events.\\
Besides its effective area, the quality of a polarimeter depends on the sampling of the Compton scattered 
events distribution (Figure \ref{KN_plot}) by the detectors. An usual figure of merit for polarimeters 
is the $Q$-factor, defined as the ratio of the parallel and perpendicular components of the scattered 
distribution. The $Q$-factor, which is energy dependent, has been calculated for SPI to be $\sim0.24$ 
over the energy range of 0.1-1 MeV \citep{2007ApJS..169...75K}.\\
When searching for polarization in a source emission, any scattered distribution on the detector 
plane will be modulated by the coded mask shadow and the dead detectors. These effects can be 
mistaken for a polarization signal if they are not properly accounted for. Instead of looking for 
an anisotropy in the scattered distribution, a much simpler method is to simulate the detector 
response for a polarized flux and compare it to the measured data. 
In this way, polarization analysis with SPI is perform in the same way than imaging but using the 
instrument response with one additional dimension: the P.A.
Thus, for each SPI observation we need simulations for different P.A.s in the source emission.

\section{SIMULATING THE SPI RESPONSE FOR A POLARIZED EMISSION}
\subsection{The Geant4 Model}
\begin{figure}
\plotone{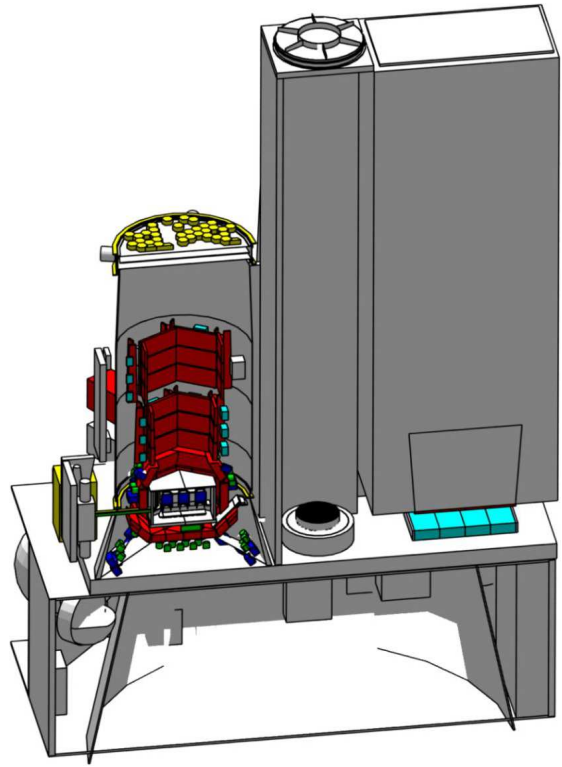}
\caption{The GEANT4 \textit{INTEGRAL} model geometry \citep{2007ClarkThesis}.}
\label{G4model}
\end{figure}

The presented simulations are based on a model recoded from the GEANT3 TIMM (for The Integral Mass Model;  
\citealt{2003A&A...411L..19F}) into the GEANT4 framework. GEANT4 \citep{2003NIMPA.506..250G} uses 
object orientated C++ language and includes extra physics such as polarization and material 
activation. 
The Geant4 software package has been tested against calibration data from a prototype detector 
array \citep{2005NIMPA.540..158M} and reliably reproducing the instrument response 
to a polarized beam. 
In the new model of \textit{INTEGRAL}, the SPI and JEM-X detectors have been both fully modeled. However, 
at the current time, the IBIS detector is only described with an approximate geometry (Figure 
\ref{G4model}). This is not a problem for our polarimetry studies, since the work revolves around the 
SPI instrument. Masses situated away from the SPI instrument will have little or no impact on the 
results. Conversely, the SPI instrument has been modeled with a high degree of accuracy. It 
comprises the active veto, the Ge detector elements inside Al capsules, surrounded by the Be 
housing, and the mask assembly. The GEANT4 incarnation of the \textit{INTEGRAL} model was specifically 
written to assess the background in the SPI instrument.
This means that several components of the geometry and particle generation processes have not been 
considered but could become important when looking at the emission from an 
astronomical object. For example, the exact geometry of the coded mask and the ability to generate 
photons with a given spectral shape had to be implemented.\\
A great deal of effort has been spent on updating the simulation to produce a much more accurate 
model of the SPI instrument, including several missing components in the geometry. 
In the current status, our simulations  
take into account the effects of non-functioning (dead) detectors, off-axis illumination and mask 
shadowing. 
Compared to the model version used by \cite{2008Sci...321.1183D}, we have included the degraded transparency 
of the central mask pixel which is 66\% at 130 keV and 80\% at 600 keV.
We have also modified the anti-coincidence system. In fact, the most important part 
surrounding the detectors was not active in the code and we introduced the anti-coincidence thresholds ($\sim$100 keV).
These modifications have an important impact on the number of events in the outer detectors 
and drastically change the instrument response.
The individual germanium crystal dimensions have 
been slightly reconsidered (resulting in a few $\%$ smaller volumes) 
modifying the instrument sensitivity.
In particular, the misalignment 
of the SPI telescope axis relatively to the \textit{INTEGRAL} satellite pointing axis has been corrected, 
which corresponds to a systematic of $\sim0^\circ.15$ in the pointing accuracy. 
This correction is essential for the simulation of the instrument response because it 
modifies the mask shadowing and thus, the event distribution over the detectors ($\sim6\%$ of the events).
These important changes in the instrument simulation lead to a good agreement with the SPI standard 
response (see Section \ref{validation}).\\
The polarimetry capability of the SPI instrument is based on the Compton efficiency of the detector 
plane, i.e., the non-diagonal terms of the response matrix ($\sim20\%$ of the total efficiency).
This Geant4 model allow us to carefully evaluate them, through their dependence on the P.A. of the incident radiation. 
It provides a robust technique, using the whole instrument response and thus 
giving the best sensitivity to the polarization characteristics.

\subsection{Simulated Data Production}
The simulated source is characterized by its direction in equatorial coordinates (R.A., decl.), an 
energy spectrum, and a P.A. This angle corresponds to the position angle of the 
 electric vector measured from north to east on the sky. 
For each pointing, simulations are carried out at 18 P.A.s (0$^\circ$-170$^\circ$ in 
$10^\circ$ steps) and 1 unpolarized. Only the P.A.s between $0^\circ$ and 
$170^\circ$ need be modeled due to the $180^\circ$ symmetry of the Compton scattering process.\\
To properly simulate the photons distribution, the data is pre-analyzed and 
fitted with a simple spectral shape. 
 This is performed with the classical SPI spectral 
analysis using the XSPEC software \citep{1996ASPC..101...17A}. The spectral models form the basis 
of a probability distribution for the fired photons so that a single photon may be simulated at 
any energy, but after millions of photons, the spectrum should resemble that of the observed 
data. In theory, it should be possible to simulate a flat input spectrum with a fixed number of 
photons in each energy bin and to normalize a posteriori the number of photons in each energy bin. 
Simulations would  need to be carried out only once and spectral changes could 
be handled in the analysis. However, this will needlessly simulate many photons at high energies 
and would drastically lengthen the running time of the simulations.\\
Besides, a small difference in the input spectrum will not lead to a big effect in the simulation. 
We have compared two simulations, with a power law slope of -3.0 and -1.5, respectively.
This results in differences less than $<10\%$ in the count rate distribution recorded in the detectors.\\
For each simulation, millions of photons are fired from the direction of the source. 
Their initial positions are randomly distributed on a $140 \times 140$ cm surface, 500 cm away from 
the detector plane, so as to illuminate all the instrument. This area has been chosen large enough to include 
the direct lighting from the source and the scattered events from the structure in the detected events. 
For a typical power law spectrum with a slope of -2.2, 50 million of photons are processed for each 
simulation resulting in $\sim500,000$ single events and $\sim50,000$ multiple events in the detector. 
For a source flux of 1 Crab observed during one pointing ($\sim$2 ks), this provides an oversampling 
of approximately 40 times.\\
For one photon fired (event), all the energy deposits inside the detectors are stored 
if no deposit occurred in the anti-coincidence system.\\
The number of simulations needed makes this part of the analysis very time consuming. One  simulation 
takes $\sim6$ hr to run on a single processor. To produce data 
corresponding to 200 SPI exposures ($\sim$400 ks), we need to run 3800  simulations ($200\times 19$). 
Using 32 10-core processors, the computing time is reduced from 950 to 3 days.

\subsection{Simulated Data Preparation}
The original data produced by Geant4 is a list of energy deposits within sensitive volumes. These 
raw data require some selection and preparation before analysis.
A low energy threshold of 20 keV is applied and the events occurring in dead detectors are removed. 
These reductions affect the multiple events: some triple events will become doubles and some doubles 
will become singles. For the data analysis, we select the double events with energy deposits occurring 
in adjacent detectors. There are 42 corresponding pseudo detectors for all the possible couples of 
adjacent detectors. Then the simulated data are energy binned and recorded in the same format as the 
SPI data to simplify the analysis procedure.\\
For a single pointing, simulations are carried out for 18 P.A.s (100$\%$ polarized) 
and 1 unpolarized. However, the analysis requires simulated data for different 
polarization percentages for each angle. These data are produced by mixing the polarized simulated distributions 
with the unpolarized one using the formula
\begin{equation}
G4(\Pi,\phi)=\frac{\Pi \times G4(\phi)}{100}+\frac{(100-\Pi) \times G4(U)}{100},
\label{Percent_equ}
\end{equation}
where $G4(\Pi,\phi)$ represents the (Geant4) simulated data for a $\Pi$ percent polarized flux at the angle $\phi$, 
$G4(\phi)$ the $100\%$ polarized simulated data at angle $\phi$, and $G4(U)$ the unpolarized simulated data.
\begin{figure}
\plotone{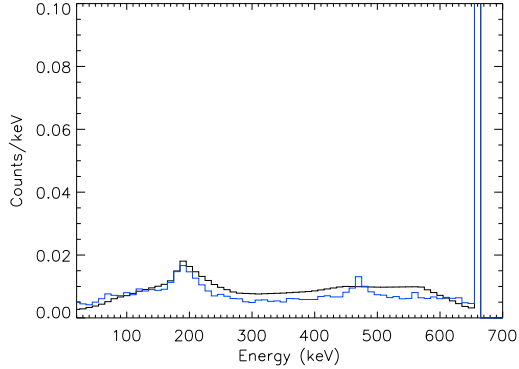}
\caption{Comparison of the SPI spectral response and GEANT4 simulation (blue) for a 
mono-energetic source (661 keV), after mass model improvements, for single events summed over all 
the detectors. The two curves, normalized to the photo-peak, differ by only $\sim8\%$.}
\label{661}
\end{figure}

\subsection{Comparison between Geant4 Model and Standard SPI Response}\label{validation}
\begin{figure}
\plotone{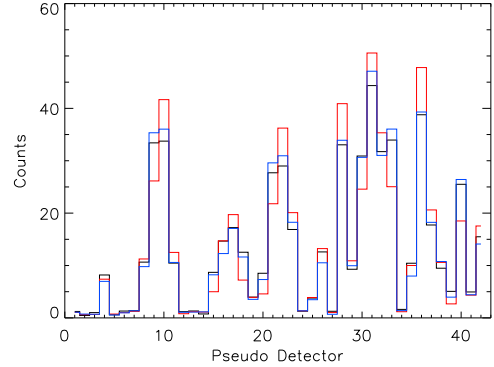}
\caption{Comparison of the SPI spatial distribution and GEANT4 simulation (blue and red) for 
the same exposure. The unpolarized simulation (blue) differs by only $\sim6\%$ and the $20^\circ$ 
polarized simulation (red) differs by $\sim20\%$.}
\label{PseudoDetCounts}
\end{figure}

The search for polarization in a source emission requires accurate simulations of the instrument 
response, any differences would lead to a spurious polarization signal.
Figure \ref{661} shows the spectral response of the GEANT4 simulation compared to the SPI standard response 
matrix \citep{2003A&A...411L..81S}, for a mono-energetic source at 661 keV. The single events are summed over all the detectors 
and the counts per 1 keV bin are plotted in fraction of the photo-peak. With the implemented mass model improvements, 
the two curves differ by only $\sim8\%$ in counts per bin with  Compton to total counts ratio 
of $0.32$ and $0.35$ for the GEANT4 simulation and the SPI response matrix, respectively. This ratio is 
important as it drives the production of multiples events. If the simulated detector geometry is 
too large or too heavy, this ratio would be smaller, i.e., less multiple events and more single events will be produced.
Figure \ref{PseudoDetCounts} shows the spatial distribution produced by the GEANT4 code compared to the 
measured SPI response, for the same configuration. Compared to the SPI response, the 
unpolarized theoretical curve (blue) differs by only $\sim6\%$ in counts per pseudo-detector whereas the 
$20^\circ$ polarized one (red) differs by $\sim20\%$. It means that the signature of a polarized signal is 
contained in the difference, i.e., $\sim15\%$ of the recorded counts. 
The spatial distribution is highly modulated by the mask pattern and the telescope pointing accuracy. 
Less obvious, the anti-coincidence system and especially its energy threshold has an important impact 
on the spatial distribution. A too high threshold will miss multiple events, increasing the recorded 
double events  in the outer pseudo-detectors.

\section{DATA ANALYSIS}
\begin{figure*}
\plotone{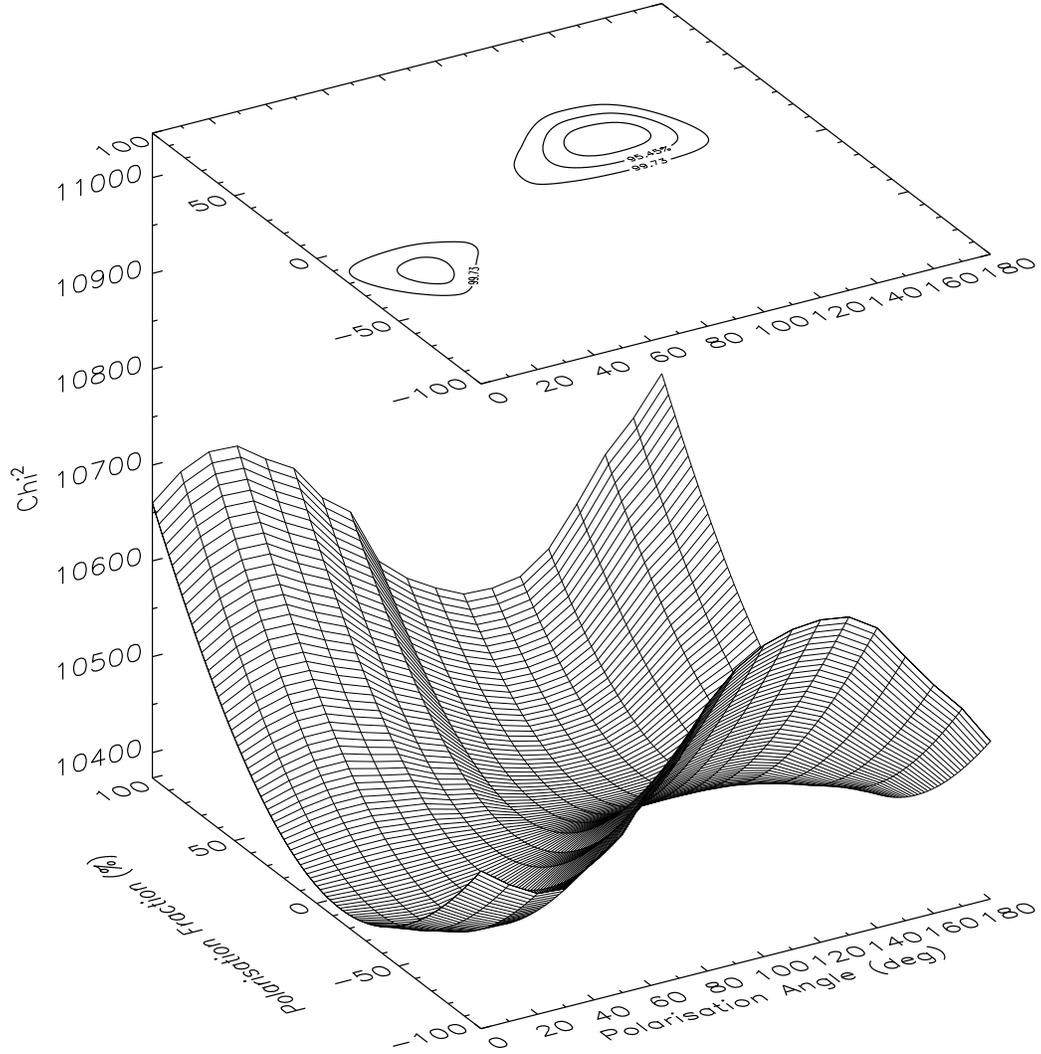}
\caption{$\chi^2$ map produced during the data analysis. There are $19 \times 201$ $\chi^2$ 
values corresponding to the best fits over the P.A.s and P.F.s. A contour plot 
is projected above showing the confidence regions ($68.27\%$, $95.45\%$, and $99.73\%$) around the 
best $\chi^2$.}
\label{chi2_map}
\end{figure*}
\subsection{Problem Formulation}
In a single SPI observation,  the recorded flux is barely enough to produce an image 
or a spectrum of the source. This means that for a more complicated analysis such as polarization 
measurement, many observations need to be summed together. Unfortunately, the instrument 
configuration changes along the time: the  pointing axis is shifted by $2^\circ$, every 30-40 minutes, 
while, on a longer timescale, detector failure must be accounted for together with
other parameter (i.e., background) evolution. 
For a given source, our study relies on simultaneous analysis of a great number of pointings.
Considering one exposure, the recorded data are modeled as
\begin{equation}
D_\mathrm{sd}=x \times G4_\mathrm{sd}(\Pi,\phi)+y \times B_\mathrm{sd},
\label{Source_equ}
\end{equation}
where $x$ is the source normalization, $G4$,  the simulated count distribution, $y$ is the background 
normalization and $B$,the background spatial distribution. The source model $G4_\mathrm{sd}(\Pi,\phi)$ 
describes the number of counts for a science window $s$, in the pseudo detector $d$ as a function of 
source polarization fraction, $\Pi$, and angle, $\phi$. The background spatial distribution is taken 
from an empty field  observation temporally close to the observation. To be comparable with the 
data, the counts issued from the simulation are renormalized to the corresponding detector 
livetimes. The $x$ and $y$ values are determined through a linear least squares resolution and the   
resulting $\chi^2$ value is stored. 
If the source and background flux can be considered constant on a long time scale, Equation 
(\ref{Source_equ}) is solved for the corresponding set of science windows.\\
Note that it is different from previous polarization studies made with the SPI instrument, 
where the source and the background were not adjusted by a fitting procedure.
For GRBs, the background was estimated using appropriate time intervals before and after 
the source signal. This is not feasible with most of the sources. 
For the Crab, \cite{2008Sci...321.1183D} considered a background with a constant pattern and 
its norm was calculated by subtracting the simulated counts from the data. This is 
hazardous because it implies the perfect knowledge of the instrument sensitivity and the source 
strength during each science window.
In our code, the background pattern is built every six months from relevant empty field observations 
and its normalization is allowed to vary on a timescale from one science window to a revolution. 
Estimating the source and the background fluxes in the analysis procedure 
introduces additional degrees of freedom (dof) but 
provides a better determination of the source contribution and 
allows polarization measurements for any (strong enough) source.

\subsection{Analysis Process}\label{process}
The analysis process relies on the comparison between the recorded data and the theoretical ones.
It can be described by the following procedure.
\begin{itemize}
\item Science windows with the source closer than $\pm 13^\circ$ from the pointing direction  
 are selected. This ensures a good illumination of the detectors.
\item A quick standard analysis is run  to filter the science window list from obvious data 
anomalies.
\item The SPI data are read into memory and the counts are summed over the selected energy range.
\item The corresponding simulated data are read into memory, the counts are summed over the 
selected energy range and weighted by the appropriate detector livetimes.
\item If the source and the background fluxes can be considered as constant, the science windows 
are grouped per data sets (with a common $x$ and $y$ parameters).
\item Looping over the entire range of P.A.s $\phi$ (0$^\circ$-170$^\circ$) and polarization fractions (P.F.s) $\Pi$
($-100\%$ to $100\%$), Equation (\ref{Source_equ}) is solved simultaneously for each data set 
through a linear least-squares resolution.
\item The corresponding $\chi^2$ are calculated, resulting in a $19 \times 201$ $\chi^2$ map in the ($\phi,\Pi$) space
(see Figure \ref{chi2_map}).
\item Using a cubic interpolation algorithm, the $\chi^2$ map is extended to $180 \times 201$.
\item The minimum $\chi^2$ is used as the indicator of the best parameters.
\item A contour plot is displayed to visualize the confidence regions at the 1$\sigma$, 2$\sigma$, and 3$\sigma$ 
levels (see Figure \ref{chi2_map}).
\end{itemize}
Some additional information are extracted from the $\chi^2$ map. 

\begin{figure}
\begin{center}
\plotone{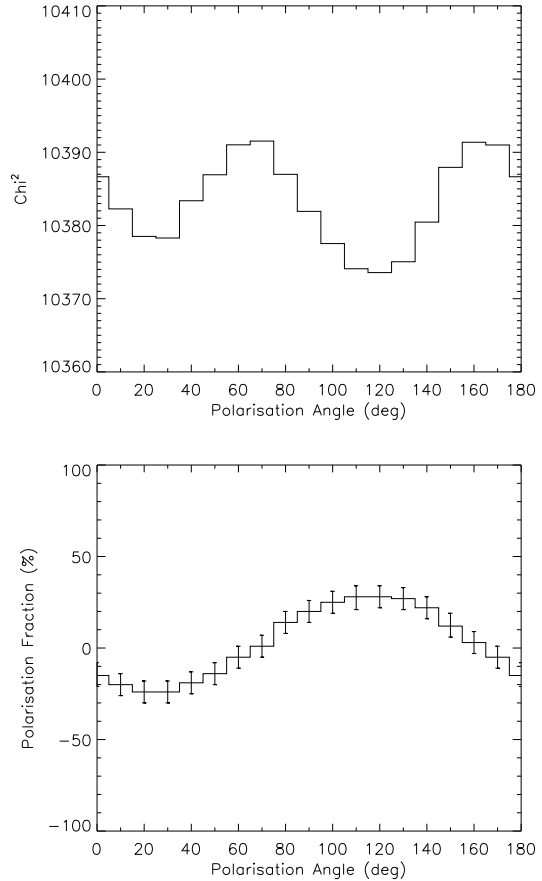}
\caption{Results extracted from the $\chi^2$ map. The P.F. giving the best fit 
for each position angle (bottom) and the corresponding $\chi^2$ (top). The errors are given at 1$\sigma$}.
\label{chi2_percent}
\end{center}
\end{figure}

Figure \ref{chi2_percent} shows the P.F. giving the best fit for each 
simulated angle (bottom) and the associated $\chi^2$ (top).

The best-fit solution appears in these curves at the 
lower $\chi^2$ value and at the maximum P.F., respectively.\\
Note that only the P.F.s from $0\%$ to $100\%$ are physically correct. 
However, the analysis is 
run over negative polarization percentages to find the best fits from a statistical point of view. 
Regarding the  angular  distribution predicted for polarized emissions (from the differential 
Compton cross-section), it is expected to have two mathematical solutions (local minima): the 
first (good) one, with a positive fraction and the second,  $90^\circ$ away, with a negative 
fraction (understandable in terms of a pattern to be subtracted from a overestimated mean flux, 
see Figure \ref{chi2_map}). This symmetry is also seen in the polarization modulation 
(see Figure \ref{chi2_percent}).

\subsection{Error Estimation}

The errors can be estimated from the $\Delta\chi^2$ method depending on the number of parameters 
of interest \citep{1976ApJ...208..177L}. 
The 1$\sigma$, 2$\sigma$, and 3$\sigma$ confidence interval for one parameter of interest, without regard to the 
value of the second, is given by $\chi^2_\mathrm{min}+1$, $\chi^2_\mathrm{min}+4$, and $\chi^2_\mathrm{min}+9$. The 
errors on the P.A. and P.F. independently are calculated with these $\Delta\chi^2$ 
values. The confidence region (two dimensions), plotted above the $\chi^2$ map, is given by the 
$\chi^2$ criteria of two parameters of interest. The $68.27\%$, $95.45\%$, and $99.73\%$ confidence 
regions are drawn at the level of $\chi^2_\mathrm{min}+2.30$, $\chi^2_\mathrm{min}+6.18$, and $\chi^2_\mathrm{min}+11.8$.\\
To be confident in our error analysis, we have verified this method by simulation using the 
Bootstrap method. 
It consists to run the same analysis thousands of 
times to quantify the errors. We apply the technique where the simulated counts are randomized 
within their (Gaussian) errors before to be re-analyzed. The best fit values of the P.A. and P.F. are stored for each trial. 
At the end, the pairs angle-fraction are spread around the real solution and form a normal 
distribution (in two dimensions). The analysis of this distribution provides the region 
(pair angle-fraction) or the interval (angle or fraction) containing $68.27\%$, $95.45\%$, and 
$99.73\%$ of the realizations. 
We then identify on the $\chi^2$ map the $\Delta\chi^2$ which gives the 
same confidence levels. 
We find that the $68.27\%$, $95.45\%$, and $99.73\%$ confidence intervals are obtained at the level
of $\chi^2_\mathrm{min}+1$, $\chi^2_\mathrm{min}+4$, and $\chi^2_\mathrm{min}+9$, and the equivalent confidence regions 
at the level of $\chi^2_\mathrm{min}+2.30$, $\chi^2_\mathrm{min}+6.18$, and $\chi^2_\mathrm{min}+11.8$ as expected.

\section{APPLICATION}
\subsection{The Crab Analysis}
\begin{figure}
\plotone{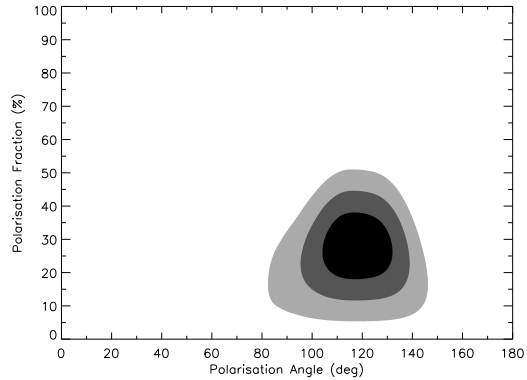}
\caption{Contour plot from the analysis of the Crab Pulsar and Nebula. The best fit ($\chi_\mathrm{red}^2=1.034$) 
indicates a P.A. of $117^\circ \pm 9^\circ$ with a P.F. of $28\% \pm 6\%$. 
The $68.27\%$, $95.45\%$, and $99.73\%$ confidence regions are shaded from dark to light gray.}
\label{Crab}
\end{figure}

The obvious candidate to search for polarization is the Crab (Pulsar and Nebula total emission) since it is one of the 
brightest source in the hard X-ray domain and the pulsar emission is expected to be highly polarized. 
Beyond that, a lot of polarization measurements exist at different wavelengths, giving a good set 
of results to compare with.\\
For this analysis, we used 239 SPI science windows from revolutions 43 to 45 (2003 February), where 
the Crab Nebula was the main target. The photons fired in the simulations follow a power law 
distribution with a spectral index of 2.2 ranging from 100 keV to 3 MeV. We have considered  
the detected counts in the energy range offering the best signal-to-noise ratio, 130 to 440 keV. 
The low energy threshold is fixed to ensure that Compton efficiency is known with enough precision. 
The high energy one is related to the source signal-to-noise ratio decrease.\\
The source and the background fluxes are considered as constant over one revolution ($\sim$3 days). 
The analysis process described in Section \ref{process} is applied, producing a $\chi^2$ map in the 
polarization angle-fraction space. Figures \ref{chi2_map} and \ref{chi2_percent}, previously quoted 
for illustration purposes, come from this Crab observation analysis. The corresponding contour plot 
with the confidence regions is highlighted in Figure \ref{Crab}. The lowest $\chi^2$ 
($\chi_\mathrm{red}^2=1.034$ with dof=10031) indicates a 
P.A. of $117^\circ \pm 9^\circ$, measured from north to east, with a P.F. of $28\% \pm 6\%$ (errors given at 1$\sigma$).\\
The P.A., which indicates the direction of the photon electric vector, appears to 
 align nicely with the spin axis of the Crab pulsar, estimated at $124^\circ \pm 0.1^\circ$ from X-rays 
imaging \citep{2004ApJ...601..479N}.
Extending the energy range to 130-650 keV, we find a P.F. of $32\% \pm 7\%$ and 
$34\% \pm 8\%$ in the energy range 130-1000 keV. We did not find any significant dependence of the 
P.F. with the energy range. This behavior is consistent with a synchrotron emission process,
which is expected for the Crab emission.
Previous polarization measurements were made on the Crab, using only the off-pulse emission 
which is expected to be much more polarized. \cite{2005AIPC..801..306K} reported a P.A. of $123^\circ$ 
in the optical domain and \cite{2008ApJ...688L..29F} a P.A. of $122^\circ \pm 7.7^\circ$ in the hard X-ray domain. 
Polarization measurements of the unpulsed emission has been already made with the SPI instrument, based 
on a previous mass model version and a different analysis process by \cite{2008Sci...321.1183D}. 
With $\sim$600 science windows ($\sim$1.5 Ms), covering the first three years of \textit{INTEGRAL} operations 
(from revolution 43 to 422 in 2006 March), they found a P.A. of $123^\circ \pm 11^\circ$ with a P.F. of 
$46\% \pm 10\%$. Our work includes only 239 science windows (570 ks) and consider the total 
(pulsar and Nebula) emission, but we find a better constrained solution, showing that the mass model 
improvements and the analysis process have strongly reduced the systematics errors.

\subsection{Systematic Errors}
To produce a spurious polarization signal requires an effect able to favor one direction
in the scattered events, i.e., affecting more counts in some specific pseudo-detectors. This kind of 
systematic error could come from the instrument itself or/and from the simulations. 
Polarimeters instruments are often designed to spin their detection plane during an observation so as to blend 
the potential detection artifacts. SPI is not spinning but its coded mask imaging capabilities require 
a specific observation strategy (dithering) during which the pointing direction shift around the source. This means that 
the detectors, so the pseudo-detectors, lit by the source are not always the same. Moreover, when using 
many observations over a one year period, the detection plane azimuthal angle (rotation around the pointing axis) 
varies by 1 deg per day (mean value).
These two characteristics drastically limit any instrumental systematic effect.\\
\cite{2007ApJS..169...75K} have investigated the potential intrinsic biases related to
the detector with ground calibration data. They could constrain them to be around 1$\%$,
quite negligible for our study.\\
On the other side, we have to check that the simulations are free of any systematic effects.
To estimate these systematics, we have applied our analysis procedure to the long distance 
imaging ground calibration data obtained at Bruy\`eres-le-Ch\^atel, France \citep{2003A&A...411L..71A}.
We have selected 10 exposures (from 12 to 38 minutes) performed with an unpolarized source of $^{137}$Cs at 125 m. 
We have produced the corresponding simulated observations with our model and selected the double events for 
which the total energy corresponds to the $^{137}$Cs line energy. Applying our analysis process we found a P.F.  
$\leq 5\%$ whatever the angle. This value is overestimated due to the uncertainties linked to the calibration 
setup such as the position and the finite distance of the radioactive source. As a second way, 
we have used the standard SPI response matrix \citep{2003A&A...411L..81S} which
provides the response for a given source position at infinity.
From this standard response matrix, we have reproduced
a set of simulated observations corresponding to the SPI observations used in our Crab analysis. 
As this standard response matrix does not include the effects of polarization in the Compton scattering distribution, 
this simulated set is representative of observations of a non-polarized source. 
We have analyzed these observations with the procedure described above and found a P.F.  
$\leq 4\%$ whatever the angle.
Consequently, this demonstrates that the imaging system of SPI (pixelated detector + coded mask + dithering) 
and the use of a complete and validated instrument response ensure reliable polarization measurements with systematic 
effects smaller than $4\%$-$5\%$.

\section{SUMMARY AND CONCLUSIONS}
We have presented specific analysis tools developed to study the polarization of emissions detected 
by the SPI spectrometer, aboard the \textit{INTEGRAL} mission, in the hard X-ray domain (130 keV to 8 MeV
for the polarimetry studies). 
Polarization information being contained in the Compton interaction parameters (diffusion and 
azimuthal angles), we used the photons which undergo a Compton interaction in one of the 19 germanium 
crystals with the diffused photons escaping to an adjacent crystal.\\
Precise simulations using Geant4 can provide the SPI instrument response for these polarized events,
for various P.A.s (from $0^\circ$ to $180^\circ$) 
and P.F.s (from $0\%$ to $100\%$). The observed data are compared with each of these 
simulations, through $\chi^2$ estimation in the same way than imaging is performed with SPI data. 
We produce a $\chi^2$ map in the two-dimensional polarization angle-fraction space and consider the lowest 
$\chi^2$ value to determine the most probable parameters of the incident flux.\\
We have first applied our method to $\sim$600 ks of observation on the Crab and found immediately 
a P.A. of $117^\circ \pm 9^\circ$, in perfect agreement with previously published value 
on the Crab \citep{2008Sci...321.1183D, 2008ApJ...688L..29F}. However, our analysis method is 
more precise than that presented in \cite{2008Sci...321.1183D} since we use an improved Geant4 version 
of the SPI/\textit{INTEGRAL} mass model and develop more sophisticated algorithms for the background determination
and source flux extraction. 
In particular, we have included the SPI alignment correction with the spacecraft axis to 
improve the  source projection into the detector plane, a parameter essential for the simulations. 
Moreover, our analysis process adjusts the source and background contributions within the fitting procedure 
and permits their variation in time.
This allows us to better determine the source contribution and to apply the polarimetry 
analysis to other sources than strong GRBs or constant sources.\\
In conclusion, the measurement of the polarization appears as a promising area in the hard X-ray 
domain, through the Compton interaction process inside detection systems. It requires a high 
statistic and thus strong sources. Nonetheless, pulsars, bright transients and GRB are as many 
potential targets for this kind of studies, which bring crucial information, very complementary 
to the spectral and timing analyzes. This promising and still unexplored field should provide 
key information on mechanisms at work inside the most powerful engines of the sky.\\

The \textit{INTEGRAL} SPI project has been completed under the responsibility and leadership of CNES.
We are grateful to ASI, CEA, CNES, DLR, ESA, INTA, NASA, and OSTC for support.
M. Chauvin and D. J. Clark gratefully acknowledge  financial support provided by the CNES.




\end{document}